\documentclass[
rsi,
 amsmath,amssymb,
reprint,%
]{revtex4-1}

\usepackage{graphicx}
\usepackage{dcolumn}
\usepackage{bm}

\usepackage[utf8]{inputenc}
\usepackage[T1]{fontenc}
\usepackage{mathptmx}
\usepackage{etoolbox}

\makeatletter
\def\@email#1#2{%
 \endgroup
 \patchcmd{\titleblock@produce}
  {\frontmatter@RRAPformat}
  {\frontmatter@RRAPformat{\produce@RRAP{*#1\href{mailto:#2}{#2}}}\frontmatter@RRAPformat}
  {}{}
}%
\makeatother
\begin{document}


\title[]{Dynamic thermal sensitivity of microwave cryogenic sapphire resonator}
\author{M.-Y. Hachani}
\author{C. Fluhr}
\author{B. Dubois}
\affiliation{ FEMTO Engineering. 15B avenue des Montboucons,  25030 Besançon}
\homepage{https://www.femto-engineering.fr/}
\author{G. Le Têtu}
\author{G. Cabodevila}
\author{V. Giordano}
\affiliation{FEMTO-ST Time \& Frequency Dpt., CNRS, UMLP, SUPMICROTECH, 26 rue de l'Epitaphe, 25000 Besançon}
\homepage{https://www.femto-st.fr/fr/Departements-de-recherche/TEMPS-FREQUENCE/Presentation}

 \email{giordano@femto-st.fr}

\date{\today}
\begin{abstract}

We have discovered a memory effect in the temperature sensitivity of a cryogenic sapphire microwave resonator, at the heart of the ultra-stable Cryogenic Sapphire Oscillators (CSOs).  Such effect is due to the relaxation time of Cr$^{3+}$ impurities, and results in hysteresis in the frequency vs temperature behavior.  These paramagnetic impurities -- always present in synthetic sapphire -- produce a temperature turning point which is necessary to achieve ultimate frequency stability.

The practical implication on the CSO is that the sapphire resonator's frequency to depends on the rate of temperature change. This dynamical thermal sensitivity results in a wide bump in the Allan deviation at $\tau\approx10 \mathrm{s}$ integration time, where the frequency stability is degraded. The actual degradation depends on the specie and on the amount of the dominant paramagnetic impurity.

\end{abstract}

\maketitle

\section{\label{sec:level1}Introduction}

The Cryogenic Sapphire Oscillator (CSO) is the microwave signal source that exhibits the lowest frequency fluctuations over measurement time up to $10^4$~s \cite{apl-2023}. Former laboratory prototypes  have demonstrated a fractional frequency stability better than $1\times 10^{-15}$ at short term \cite{rsi-2012,hartnett-2012-apl}. 
Autonomous and reliable semi-commercial units are currently available with a conservative Allan Deviation (ADEV) specification such as $\sigma_y(\tau) \leq 3\times 10^{-15}$ for $1~$s$\leq \tau \leq 10,000$~s and better than $1\times 10^{-14}$ over one day \cite{im-2023}. 
The exceptional performance of the CSO primarily results from the intrinsic properties of its frequency reference: the microwave cryogenic whispering-gallery mode sapphire resonator. The resonator, machined from a high-purity Al$_2$O$_3$ mono-crystal, is cooled to a few degrees above the liquid helium temperature, typically between 5 and 8 K. At its operating temperature, the resonator presents a Q-factor of one billion typically and its temperature sensitivity is minimised. Yet, despite these excellent performances, the frequency stability of the CSO remains limited by technical issues that are still poorly understood. In particular, we have already experimentally demonstrated that residual temperature fluctuations degrade the short-term frequency stability beyond the expected limit, considering the thermal sensitivity of the resonator and the noise floor of the temperature controller~\cite{uffc-2016}.\\
In this paper, we experimentally highlight a dynamic sensitivity of the resonator that had never been observed before. This dynamic sensitivity, which causes the sapphire resonator's frequency to depend on the rate of temperature change, helps to explain the observed frequency stability limitations and gives clues to overcome this problem. To understand this dynamic sensitivity, we involve the relaxation time of paramagnetic impurities present in the crystal and responsible for first order thermal compensation.

\section{Low consumption CSO present status}
\begin{figure}[h]
\centering
\includegraphics[width=0.5\textwidth]{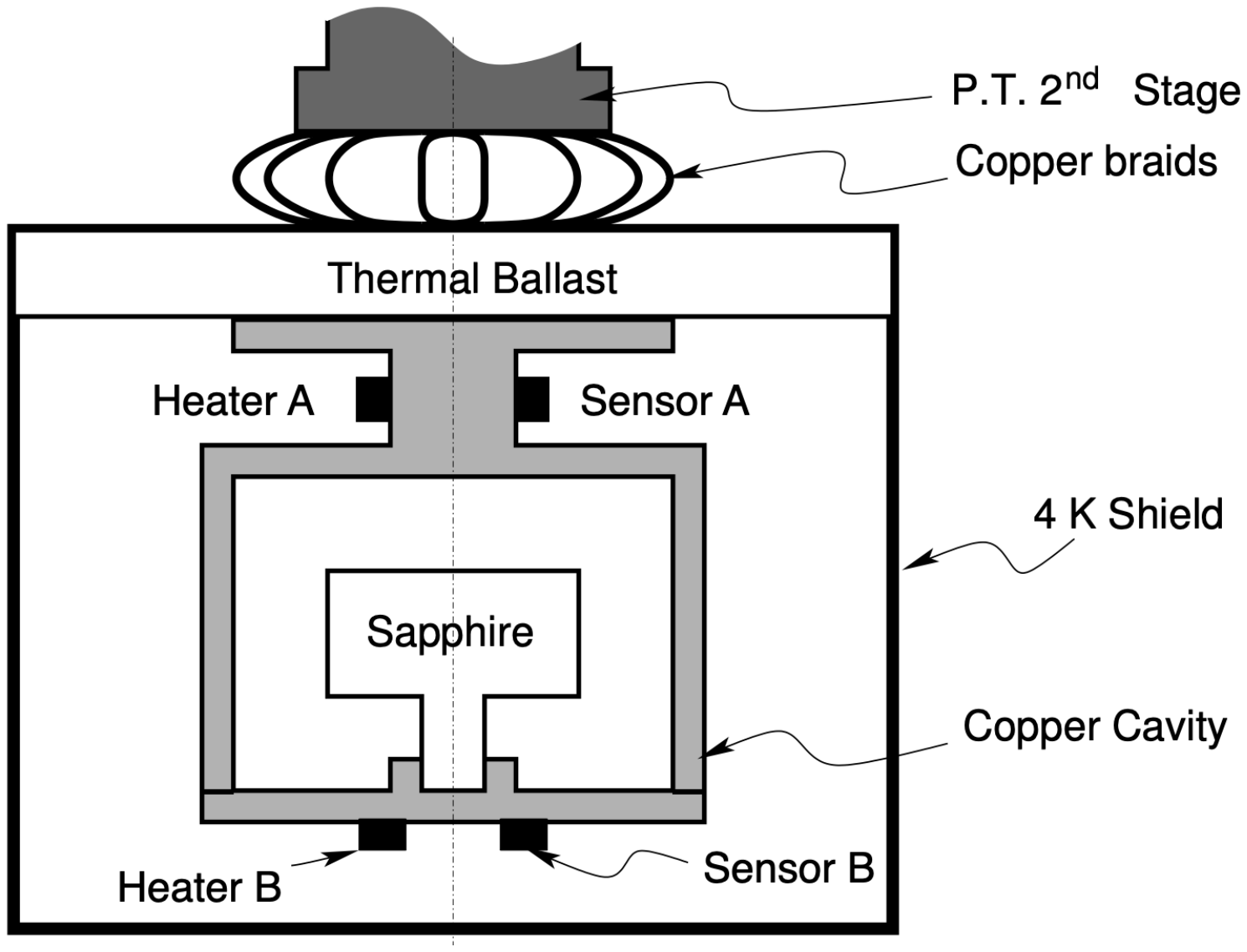}
\caption{\small \it The ULISS-2G resonator cooled using a low consumption Pulse-Tube (P.T.) cryocooler}
\label{fig-1}
\end{figure}

The low consumption CSO technology issued from the Femto-ST Institute evolved into a semi-commercial product code named ULISS-2G, available from FEMTO- Engineering  
to qualified users \cite{cryogenics-2016,www.uliss}. To date, seven units have been deployed in national metrology institutes (NMIs) worldwide, where they primarily operate as local oscillators for atomic fountain primary frequency standards.\\
The CSO is a Pound–Galani oscillator whose reference frequency is provided by a cryogenic microwave sapphire resonator schematized in the figure \ref{fig-1}.

The ULISS-2G resonator is a cylinder of sapphire mono-crystal of 54 mm diameter and 30 mm high, resonating in quasi-transverse magnetic whispering-gallery mode WGH$_{15,0,0}$ at $\nu_0 =9.99~$GHz $\pm~5~$MHz. A spindle is machined from the bulk together with the resonator. The resonator is placed in the center of a cylindrical gold-platted copper cavity. A clamp with one of the jaws attached to the cavity bottom flange rigidly maintains the resonator with no stress in the resonator circumferential region, where the microwave energy is located. This assembly is housed in a compact cryostat (volume of 34 dm$^3$) and cooled by a pulse-tube cryocooler providing $250$~mW of cooling power at $4.2$~K. Each resonator exhibits a characteristic turnover temperature $T_0$, typically between 5 K to 8 K, at which its first-order sensitivity to temperature fluctuations vanishes. In the current ULISS-2G configuration, the resonator temperature stabilization is carried out using a commercial temperature controller (LakeShore 350) associated with a high sensitivity Cernox thermal probe (sensor A) and a resistive heater (heater A) placed on the post linking the cavity to the thermal ballast.   In the experiments described in this article, we placed a second sensor/heater pair (noted B in Fig. \ref{fig-1}) on the cavity bottom flange, near the resonator maintaining clamp.\\

Within the bandwidth of the Pound servo loop, the ultra-high frequency stability of the sapphire resonator is transferred to the oscillator output signal. In principle, the achievable frequency stability is ultimately limited by the detector white noise from which the Pound error signal is derived \cite{rsi-2016}. 
This limitation depends on the resonator tuning parameters, including its loaded Q-factor, coupling coefficients, and the injected power. For a properly optimised resonator, the achievable fractional frequency stability is better than $7\times 10^{-16}\tau^{-1/2}$ where $\tau$ is the integration time.\\

\begin{figure}[h]
\centering
\includegraphics[width=0.5\textwidth]{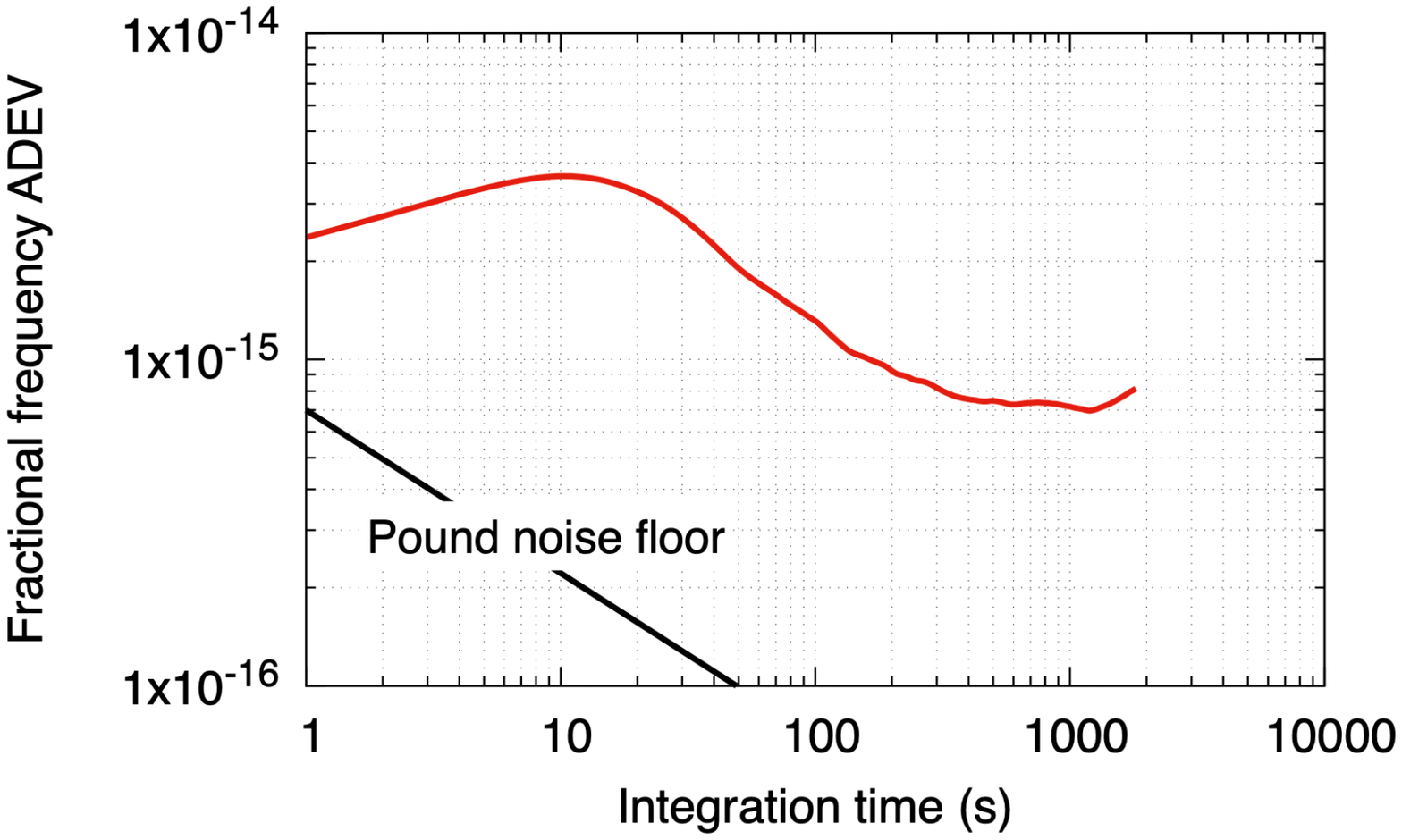}
\caption{\small \it U12 prototype fractional frequency stability.}
\label{fig-2}
\end{figure}

 Figure \ref{fig-2} shows the Allan deviation of U12: the latest ULISS-2G prototype under development.

\begin{figure}[h]
\centering
\includegraphics[width=0.5\textwidth]{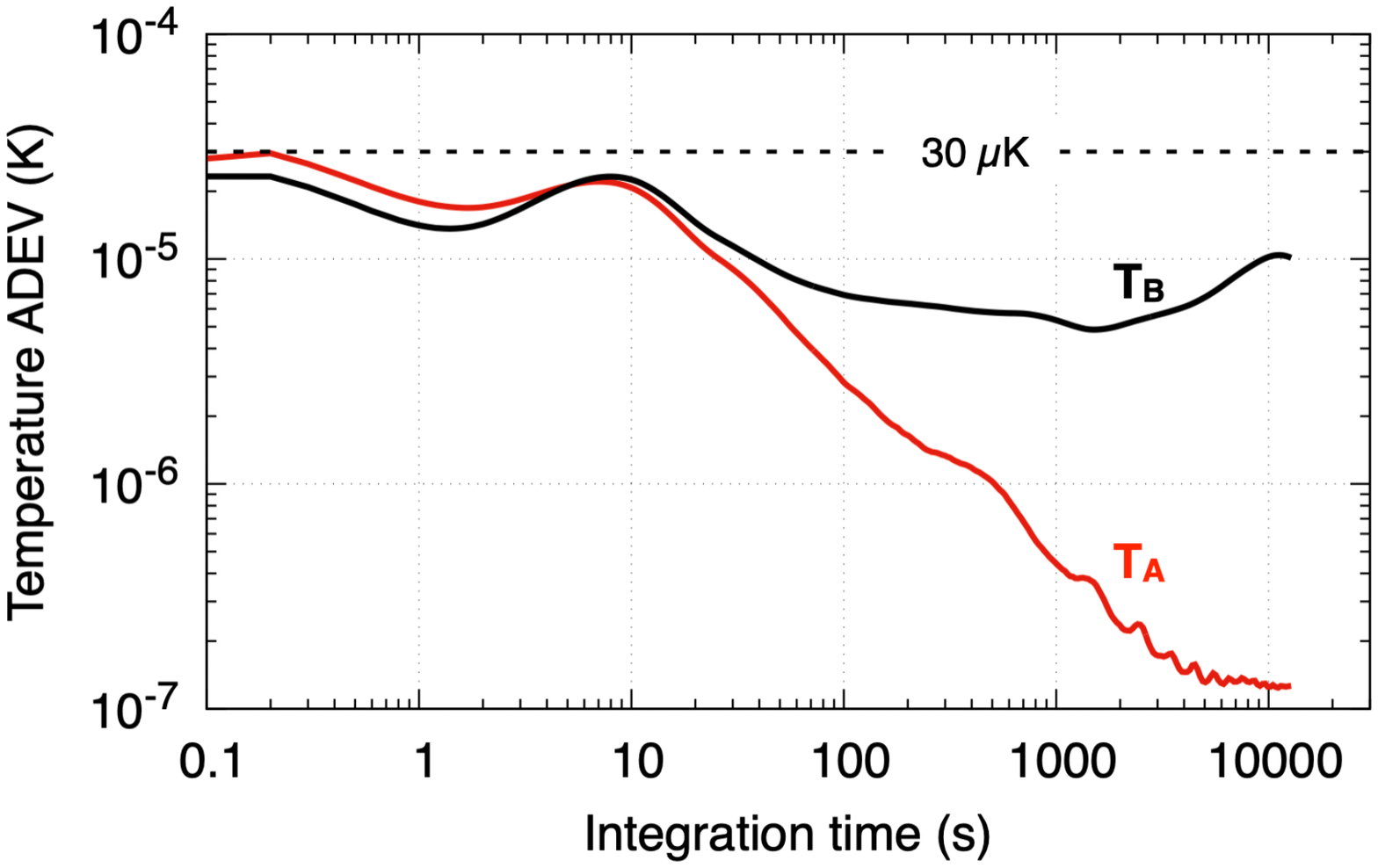}
\caption{\small \it Temperatures stability in ULISS-2G.}
\label{fig-3}
\end{figure}

As pointed out in \cite{giordano2013uliss,uffc-2016}, the bump observed in the ADEV around $10~$s results from residual temperature fluctuations. In the ULISS-2G configuration, the resonator is subjected to temperature variations originating from the cryocooler and from thermal exchanges with the ambient temperature. A stainless steel thermal ballast is used to passively filter the 1.4 Hz temperature modulation generated by the cryocooler. However its efficiency drops significantly at lower frequencies. Increasing its mass without damaging the internal mechanical suspensions and  without compromising the ULISS-2G's integrable and compact design is challenging.
Rejecting the temperature disturbances therefore requires a wide bandwidth for the temperature control. However, the delays and phase shifts inherent to the thermal assembly limit the achievable bandwidth typically between 0.1 Hz and 1 Hz. We are therefore constrained to find a compromise between the rejection of external perturbations and the stability of the system. The bump in the ADEV results from the slight amplification of temperature fluctuations near the frequency where the phase margin is zero. The time position and amplitude of the ADEV degradation depend on the Proportional-Integral-Derivative (PID) controller tuning, although it cannot be completely eliminated. Despite these difficulties, it is possible, after careful adjustment of the controller parameters, to achieve a temperature stability better than $30~\mu$K as demonstrated in the figure \ref{fig-3}.

For the sensor A used in the control loop, at short integration times ($\tau < 1$~s), the temperature ADEV is limited by the intrinsic noise of the temperature probe, which results from the limited resolution of the temperature measurement. Then, we observe near $10~$s, the bump revealing the control bandwidth and eventually an improvement of the temperature stability with the integration time as expected. The temperature $T_B$ presents higher fluctuations but nevertheless its ADEV remains below $30~\mu$K even for long integration time.
Taking into account the thermal sensitivity generally assumed for the resonator, such temperature fluctuations should not limit the oscillator stability above a few $10^{-16}$, which is an order of magnitude below the experimentally observed stability. The remainder of this article attempts to address this apparent contradiction.

\section{Static thermal sensitivity}
The presence of paramagnetic dopants in the sapphire crystal induces the appearance of a turnover temperature $T_0$ in the frequency-temperature dependance \cite{bun1986,giles1990,mtt-2015}.
Indeed, high-purity sapphire crystals invariably contain a small concentration of paramagnetic impurities, such as Cr$^{3+}$ or Mo$^{3+}$. At low temperatures, these residual impurities give rise to a temperature-dependent magnetic susceptibility, which compensates for the intrinsic thermal sensitivity of the resonator. Thus, the resonator frequency-temperature dependence can be written as \cite{mann92-jpd}:

\begin{equation}
\dfrac{\nu(T)-\nu_{0}}{\nu_{0}} = A T^{4}+ \frac{\chi(T)}{2} =  A T^{4}+ \frac{C}{T}
\label{equ-1}
\end{equation}  
The first term proportional to $T^4$, combines the temperature dependence of the dielectric constant and the thermal expansion \cite{luiten96}. It is the intrinsic temperature sensitivity of an hypothetic resonator made in a perfect sapphire mono-crystal without any impurity, whose resonance frequency at $T=0~$K is $\nu_{0}$. The parameter $A$ is slightly mode-dependent and is equal to $ -2.95\times 10^{-12}$ K$^{-4}$  for the  WGH$_{15,0,0}$ mode we are using. The second term represents the contribution of the paramagnetic dopants diluted with a small concentration in the actual sapphire resonator.  
$\chi(T)$ is the real part of the susceptibility induced by these paramagnetic dopants. In the temperature range we are experienced, the susceptibility follows the Curie law and thus is inversely proportional to the temperature. The parameter $C$ depends on the species and concentration of paramagnetic impurities. For modes lying under the Electron Spin Resonance (ESR) frequency of the paramagnetic ion, $C$ is negative and the derivative of expression \eqref{equ-1} nulls at a given temperature $T_0=\sqrt[5]{C/4A}$.

\begin{figure}[h]
\centering
\includegraphics[width=0.5\textwidth]{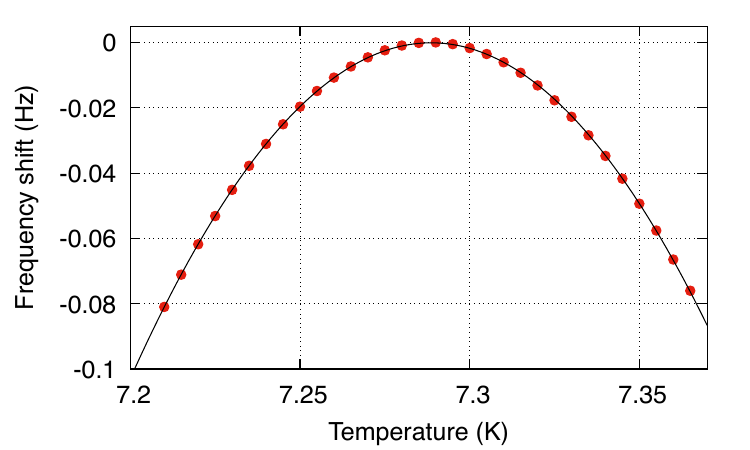}
\caption{\small \it U12 ~frequency vs temperature around $T_0 \approx 7.288~$K.}
\label{fig-4}
\end{figure}

This thermal compensation is essential for achieving the highest frequency stability as, at $T_0$, the resonator frequency thermal sensitivity is minimised. Figure \ref{fig-4} shows the U12 oscillator frequency variation as a function of the resonator temperature between $7.2$~K and $7.37$~K. 

To draw this characteristic, the frequency of the beat-note between the output of the oscillator under test and the one of another CSO is recorded using a high resolution frequency counter. The resonator temperature set-point is changed step by step. Then, the beat-note frequency is collected at each step and after waiting for the complete stabilisation of the system. At least 2 hours are needed to draw such a thermal characteristic.  Assuming $T=T_0+\Delta T$ with $\Delta T/T_0\ll 1$ and expanding the expressions involving $T$ to the second order in $\Delta T/T_0$, the expression \eqref{equ-1} reduces, near $T_0$, to a second-order function :

\begin{equation}
\frac{\nu(T)-\nu(T_0)}{\nu_0} = a  (T-T_0)^2~~~\mathrm{with~} a = \dfrac{5}{2}\dfrac{C}{T_0^3}
\label{equ-2}
\end{equation}

where $a$ is the curvature of the frequency to temperature characteristic.  For the U12 resonator, we have $T_0 =7.288~$K, $\nu_0=9.998~$ GHz and $a=-1.3\times 10^{-9}~\mathrm{K}^{-2}$. Thus, by differentiating the previous expression, the relative frequency fluctuations related to the temperature variations $\delta T$ are:

\begin{equation}
\frac{\delta \nu}{\nu_0} = 2a (T-T_0)\delta T
\label{equ-3}
\end{equation}

Owing to the high frequency stability of our CSOs, the turnover temperature can be determined within $\pm 2~$mK, yielding $T-T_0 \leq 2~$mK. A temperature fluctuation of $\delta T \approx 30~\mu$K should therefore limit the CSO fractional frequency stability to $1.5\times 10^{-16}$, which is one order of magnitude below the experimental observation. This discrepancy questions the validity of Eq. \eqref{equ-2}.

\section{Dynamic thermal sensitivity}
The present situation is similar to that reported more than 40 years ago for quartz oscillators, where the observed frequency instabilities were several orders of magnitude higher than the limits predicted from the resonator’s static thermal sensitivity and the known temperature fluctuations of the oven. To account for the experimental observations, a phenomenological term proportional to the time derivative of the temperature was introduced into the frequency–temperature law of quartz resonators \cite{ballato1979,gagnepain1989}. This hypothesis was later confirmed by calculating the time dependent spatial temperature distribution in the quartz resonator submitted to temperature transients \cite{valentin1984,shmaliy1999}. We follow here the same approach and rewritte Eq.~\eqref{equ-2} as:

\begin{equation}
\frac{\nu(T)-\nu(T_0)}{\nu_0} = a (T-T_0)^2 +\tilde{a}\dfrac{dT}{dt}
\label{equ-4}
\end{equation}

To test the validity of this equation and determine the value of the $\tilde{a}$ term we conducted several experiments by varying the resonator temperature.

\subsection{Temperature ramping}

\begin{figure}[h]
\centering
\includegraphics[width=0.5\textwidth]{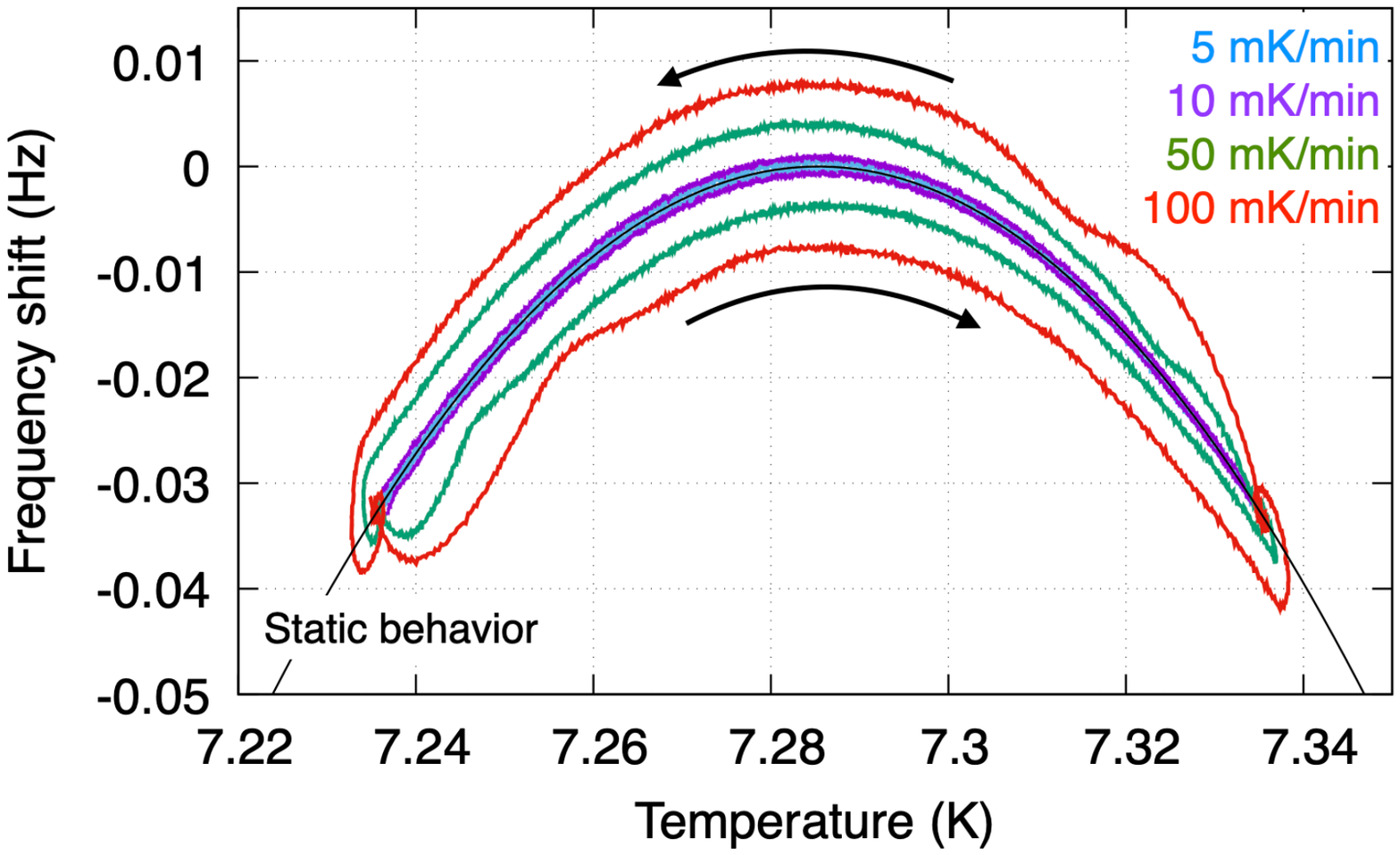}
\caption{\small \it Frequency hysteresis arising from temperature ramping.}
\label{fig-5}
\end{figure}

By using the set-point ramping functionality of our temperature controller, the resonator temperature can be ramped with a rate that can be varied. 
Figure \ref{fig-5} shows the frequency variation when the temperature is ramped up and down with several rate values.

As soon as the temperature ramp is started, the frequency comes out of the static characteristic.
The frequency follows a counter-clock-wise hysteresis curve, meaning that $\tilde{a}$ is negative. As the rate of temperature change increases, the deviation from the static behavior becomes larger as expected from Eq. \eqref{equ-4}. For a temperature ramp, the rate is constant and near $T\approx T_0$ the frequency shift is simply $\tilde{a} \frac{dT}{dt}$. From the experimental variation, we deduced:%
\begin{equation}
\tilde{a} \approx - 4.5\times 10^{-10}~\mathrm{sK}^{-1}
\label{equ-5}
\end{equation}

 \subsection{Sinusoidal temperature modulation}
 
\begin{figure}[h]
\centering
\includegraphics[width=0.5\textwidth]{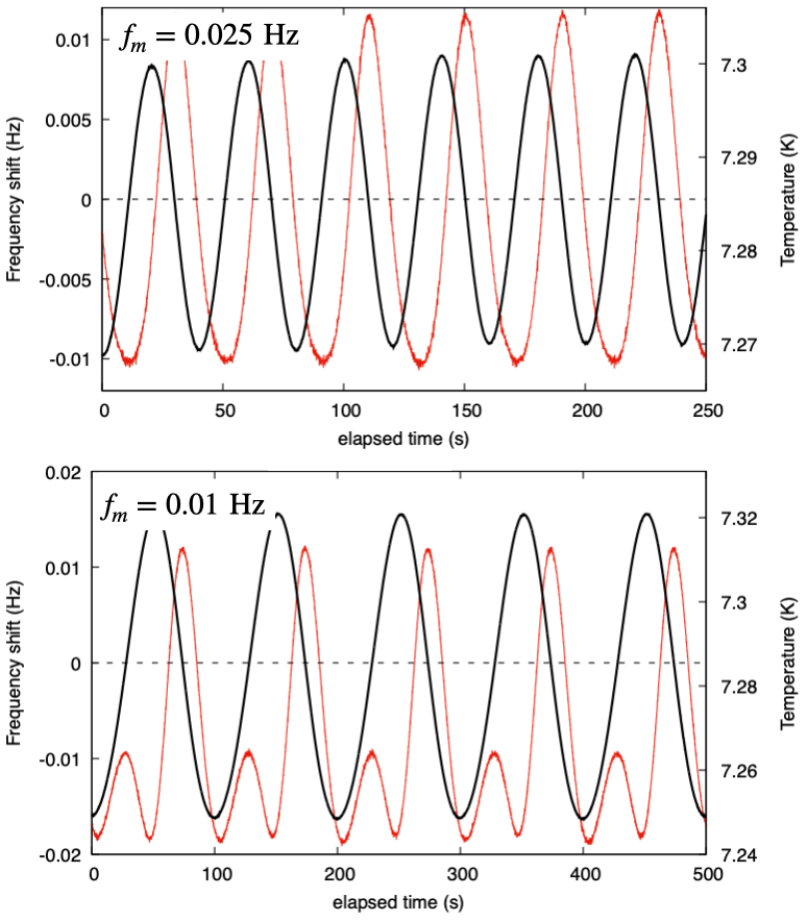}
\caption{\small \it U12~ frequency and resonator temperature. The temperature is sinusoidally modulated around $T_0 \approx 7.288~$K with $f_m=0.025$~Hz and $f_m=0.01$~Hz.}
\label{fig-6}
\end{figure}

A DC current with an added small modulation is sent  to the heater-$B$. By adjusting the set-point fixing $T_A$, the DC current and the modulation amplitude, it is easy to obtain quasi-sinusoïdal modulation of $T_B$ around $T_0$: $T_B(t) \approx T_0+\Delta T_m \sin{2 \pi f_m t}$. 
 A modulation amplitude $\Delta T_m$ of few tenths of degree can be obtained without altering the harmonic contains of $T_B$. Figure \ref{fig-6} shows the temperature $T_B$ (black line) and the CSO frequency (red line) for two values of $f_m$, i.e. $0.025$~Hz and $0.01$~Hz. 
  
Note that for $f_m=0.025$~Hz, the frequency of the CSO remains modulated at the same rate, whereas a doubling of the modulation frequency is expected from Eq. \eqref{equ-1} when $T \sim T_0$. The second harmonic appears for $f_m=0.01$~Hz as the two terms in the right hand side of Eq. (3) are of the same order of magnitude. Note also, for $f_m=0.025$~Hz the $90^{\circ}$ phase shift between the two curves, compatible with a CSO frequency following the temperature derivative. \\

\begin{figure}[h]
\centering
\includegraphics[width=0.5\textwidth]{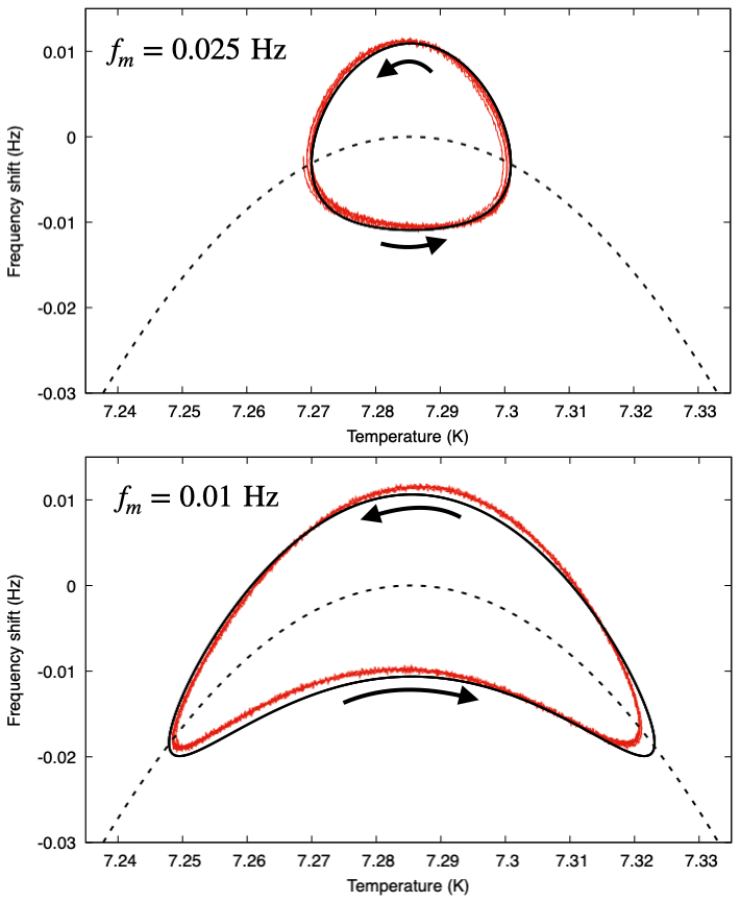}
\caption{\small \it U12~ frequency vs resonator temperature. The temperature is sinusoidally modulated around $T_0 \approx 7.288~$Kwith $f_m=0.025$~Hz and $f_m=0.01$~Hz.}
\label{fig-7}
\end{figure}

Combining the temperature and frequency data, we draw the hysteresis curves represented in the figure \ref{fig-7}.

The experimental data are represented by the red line. The black line curves were calculated from Eq. \eqref{equ-4}, with $\tilde{a}=$~$-4.5\times 10^{-10}$~sK$^{-1}$.

\section{Cause of  the dynamical thermal sensitivity}

The term $\tilde{a}$ was introduced phenomenologically into the equation describing the resonator frequency as a function of temperature. We will show now that this parameter can be related to a physical phenomenon, and we will calculate a theoretical value that is very close to the experimental value given by equation \eqref{equ-5}.\\

\subsection{Thermal gradient}
\begin{figure}[h]
\centering
\includegraphics[width=0.5\textwidth]{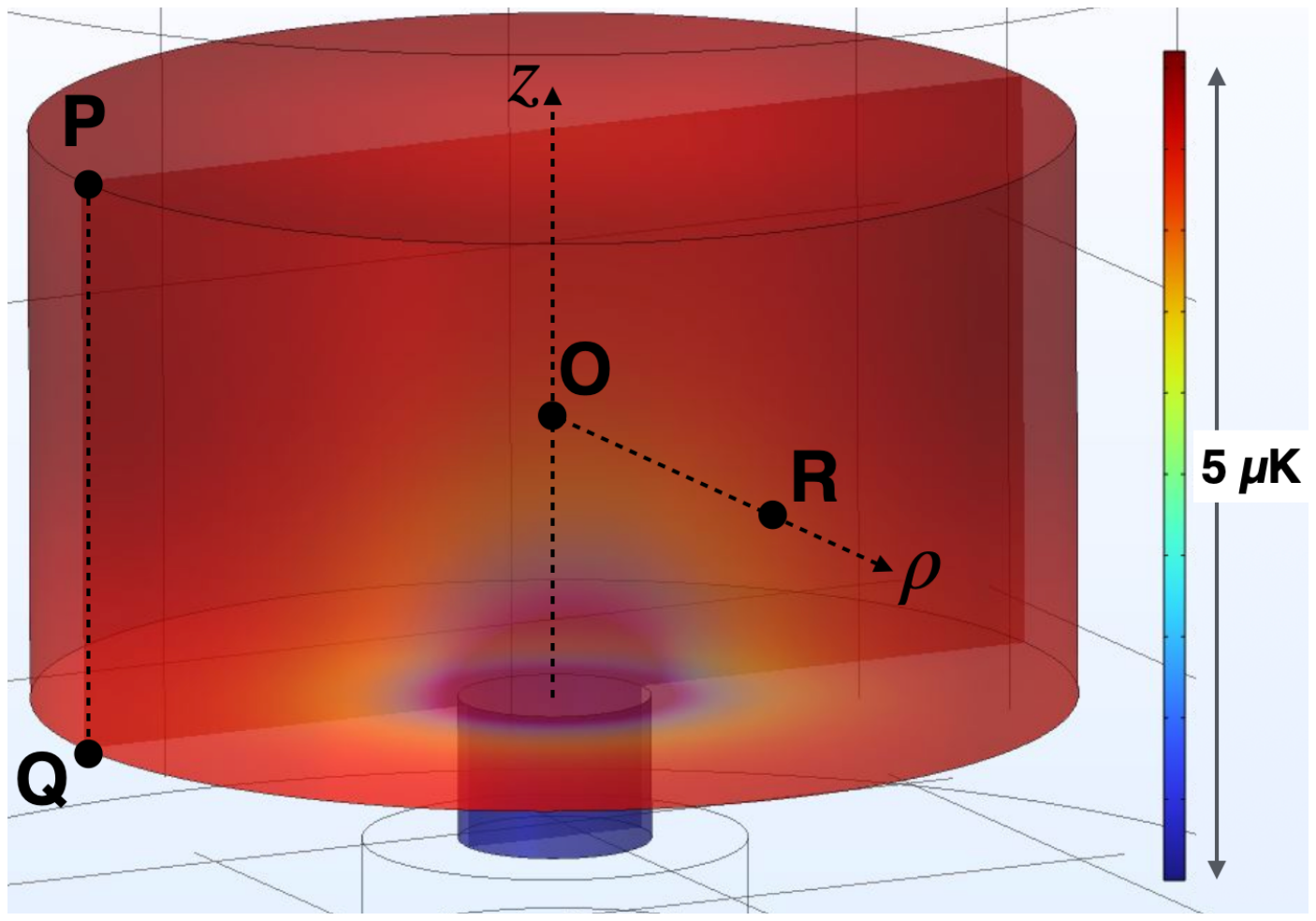}
\caption{\small \it Simulated temperature in the sapphire resonator. The mean temperature is $T\approx 7.3$~K and the full colour scale is only $5~\mu$K.}
\label{fig-8}
\end{figure}

For quartz resonators, the term  $\tilde{a}$ has been attributed to temperature gradients within the crystal. These gradients evolve as a function of the rate of change of the imposed temperature, thereby giving rise to a dependence of the resonator frequency on the time derivative of temperature. However, it is known that sapphire is the material with the highest thermal diffusivity at low temperatures \cite{ekin2006}. Following a thermal perturbation, the temperature of the crystal therefore homogenises very rapidly, which contradicts the previous hypothesis. To confirm this analysis, we performed finite-element (FE) simulations using COMSOL, which show that under the conditions of the experiments described in Section IV, the temperature gradients remain very small. The figure \ref{fig-8} shows the simulated resonator. 

We included the internal heat generation resulting from the dissipated electromagnetic power (typically of $100~\mu$W). The base of the resonator spindle is submitted to a modulated temperature around $7.3$~K with an amplitude of $15~$mK and a period of $10$~s.  We observe temperature gradients evolving over time at the same rythm that the applied modulation. Thus, the temperature difference, i.e. $T_O-T_R$, between the center of the resonator and its periphery (radial gradient) is modulated with a pic-to-pic variation of $1.4~\mu$K, while for the axial gradient ($T_P-T_Q$), the variation is about $1~\mu$K. Assuming these gradients and a temperature varying linearly with the coordinates $r$ et $z$, we calculated the resulting frequency shift using a first-order electromagnetic perturbational method \cite{[Section 6.7~]pozar2011microwave}. We obtained a relative frequency variation smaller than $10^{-16}$, which is insufficient to account for the experimental observations.

\subsection{Relaxation times of paramagnetic impurities}

For our resonator, thermal compensation arises primarily from Cr$^{3+}$ ions, which are the dominant impurities, with a concentration on the order of 60 ppb. 
The relationship given in Eq. \eqref{equ-1} holds only under constant-temperature conditions. The induced magnetic susceptibility is proportional to the population difference between the two ground states of the Cr$^{3+}$ ion, separated by an energy $h \nu_{ESR}$, with $\nu_{ESR}=11.4~$GHz. At thermal equilibrium, the populations of these two levels follow the Boltzmann distribution. Following a temperature change, the populations evolve toward the new equilibrium, with a characteristic relaxation time $\tau_1$. For high quality sapphire crystal and low Cr$^{3+}$ concentration, the spin-lattice relaxation time $\tau_1$ is $\sim 100$~ms at $7.3$~K \cite{standley65,bates1969,sewani2020}.  If the temperature variations are not too rapid, everything behaves as if the ions experience the crystal’s temperature with a delay $\tau_1$. Now, assuming the temperature is changing, we should write Eq. \eqref{equ-1} as:

\begin{equation}
\dfrac{\nu(T(t)) -\nu_0}{\nu_0} = A T^{4}(t)+ \frac{C}{T(t-\tau_1)}
\label{equ-6}
\end{equation}

For  small temperature variations, using the Taylor expansion to the first order, we have:

\begin{equation}
T(t-\tau_1) \approx T(t)-\tau_1 \dfrac{dT(t)}{dt} +\mathcal{O}(\tau_1^2) 
\label{equ-7}
\end{equation}

Making the usual approximation and for greater clarity, omitting the explicit time dependence of the temperature, we have:

\begin{equation}
\dfrac{\nu(T) -\nu_0}{\nu_0}  = A T^{4}+ \frac{C}{T} + \dfrac{C\tau_1}{T^2}\dfrac{dT}{dt}
\label{equ-8}
\end{equation}

The two first terms in the right-hand side represent the static sensitivity, which nulls for $T= T_0$. The last term proportional the time derivative of $T$ gives the expected value of $\tilde{a}$:

\begin{equation}
\tilde{a} = \dfrac{C\tau_1}{T^2_0} \approx -4.6 \times 10^{-10}~\mathrm{sK}^{-1}
\label{equ-9}
\end{equation}

This theoretical value is very near the experimental one, i.e. $ -4.5 \times 10^{-10}~\mathrm{sK}^{-1}$, which reinforces our belief that the observed effect results from the time it takes for the Cr$^{3+}$ ions to return to thermodynamic equilibrium.

\section{Frequency stability limitation}
In this section, we show that the previously highlighted dynamic sensitivity makes possible the explanation of the degradation of the CSO’s frequency stability. However, our simple model must first be refined in order to account for small random fluctuations around the inversion temperature. Indeed, the approximation that models the thermal behavior of the ions as a simple delay in the case of relatively slow and deterministic variations is no longer valid for random temperature fluctuations. In particular, equation \eqref{equ-7} no longer makes any sense, since the temperatures at times separated by a delay $\tau_1$ are uncorrelated.

\subsection{Frequency domain analysis}

Let us return to Eq. \eqref{equ-1}, which gives the frequency variation as a function of temperature. We are interested in small deviations around the inversion temperature, i.e., $T(t)=T_0+\delta T(t)$. The two terms on the right hand side of Eq. \eqref{equ-1} exhibit different dynamics in response to a temperature variation:
 
i) the first one proportional to $T^4$ represents the lattice contribution to the frequency variation. For the time scales we are considering here, it reacts quasi instantaneously. Indeed, due to the high diffusivity of the sapphire at low temperature, i.e., $D\sim 5$~m$^2/$s at $7$~K, the thermal waves propagate very rapidly inside the resonator. For a solid with a characteristic dimension  $L$, the thermal relaxation time is $\tau_{th}=L^2/(\pi^2 D)$ \cite{hahn2012heat}. With $L\sim 3$~cm and sapphire at $7~$K, we have $\tau_{th} \ll 1~$ms. We thus assume that the lattice contribution follows without any delay the temperature perturbation. Making the usual approximations, the Fourier transform of the lattice contribution to the frequency variation is:
\begin{equation}
\dfrac{\delta \nu}{\nu}(f)\vert_{lattice} = 4AT_0^3 \delta T(f)
\label{equ-11}
\end{equation}
 
 ii) the second term represents the paramagnetic impurities contribution. The dynamic of the magnetic susceptibility is described by the differential equation \cite{siegman_maser}:
  \begin{eqnarray}
\tau_1 \dfrac{d \chi(t)}{dt}+\chi(t)	& =  &\chi^{eq}(T(t))  \nonumber,\\
			& \approx  &\chi^{eq}(T_0) + \dfrac{d\chi^{eq}}{dT} \vert_{T_0} \delta T(t)
			\label{equ-12}
\end{eqnarray}

 $\chi^{eq}$ be the susceptibility at the thermal equilibrium, which depends on the actual temperature. This equation is identical to the one describing the response to a first order filter with a time constant $\tau_1$, and thus in the frequency domain we have:

 \begin{equation}
\dfrac{\delta \nu}{\nu}(f)\vert_{para} = -\dfrac{C}{T_0^2} \dfrac {1}{1+j2 \pi f \tau_1} \delta T(f)
\label{equ-13}
\end{equation}
 
Combining equations \eqref{equ-11} and \eqref{equ-13}, we can write the spectrum of the fractional frequency fluctuations $y=\frac{\delta \nu}{\nu}$ as:
\begin{equation}
S_{y} (f)=  \tilde{a}^2 \left ( \dfrac{4 \pi^2 f^2 }{1+4 \pi^2 f^2 \tau_1^2} \right ) S_{\delta T}(f)
\end{equation}
This expression shows us how the temperature fluctuations affect the oscillator frequency through the dynamical thermal sensitivity of the sapphire resonator.
Knowing the spectrum of the temperature fluctuations, it permits to evaluate the spectrum of the fractional frequency fluctuations and thus the frequency stability (ADEV) \cite{rubiola-2023-enrico-chart}.
One example is given in the next section.

\subsection{Actual temperature fluctuations and ADEV limitation}

\begin{figure}[h]
\centering
\includegraphics[width=0.5\textwidth]{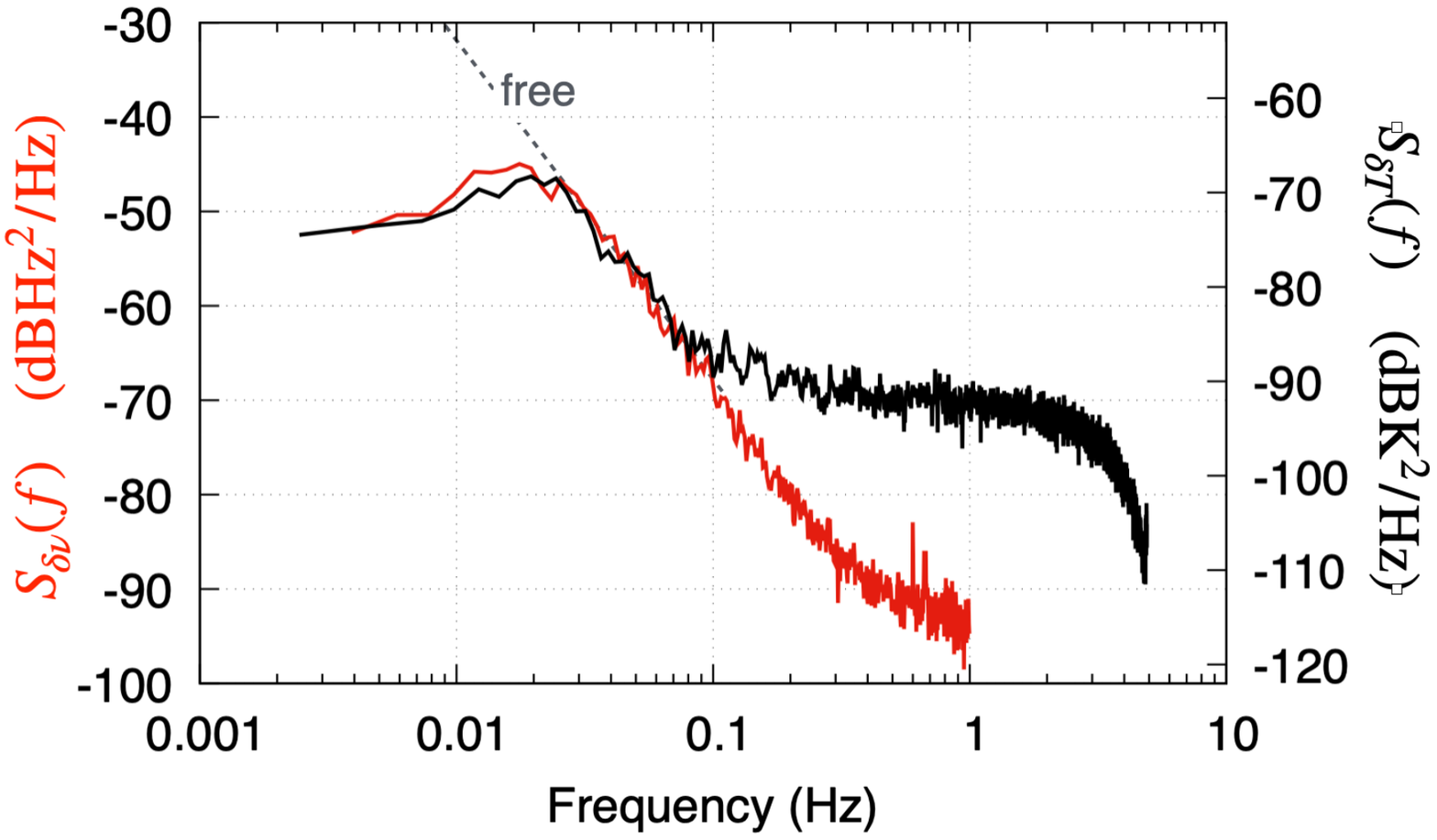}
\caption{\small \it  $S_{\delta \nu}(f)$ (red line) and $S_{\delta T}(f)$ (black line) measured at $T_0+500$~mK. The black dotted line represents the temperature fluctuations when the control is off.}
\label{fig-9}
\end{figure}

Figure \ref{fig-3} presented the temperature fluctuations as revealed by the thermal probe, which can differ from the actual resonator temperature variations due to the limited measurement resolution.
To evaluate the actual resonator temperature fluctuations, we stabilize its mean temperature at $T_1=T_0+500~$mK. In this regime, the static thermal effect dominates, and equation \eqref{equ-3} applies. For a 500 mK offset from the inversion temperature, the slope of the frequency–temperature curve is $13~$Hz/K yielding:  $S_{\delta \nu}(\omega)=\nu_0^2 S_y(\omega)=S_{\delta T}(\omega)+22~$dB. Figure \ref{fig-9} shows the power spectral density (PSD) of the frequency fluctuations, compared to the PSD of the temperature fluctuations measured by the thermal sensor attached to the flange supporting the sapphire resonator. The two vertical scales have been adjusted to account for the 22 dB difference.

\begin{figure}[h]
\centering
\includegraphics[width=0.5\textwidth]{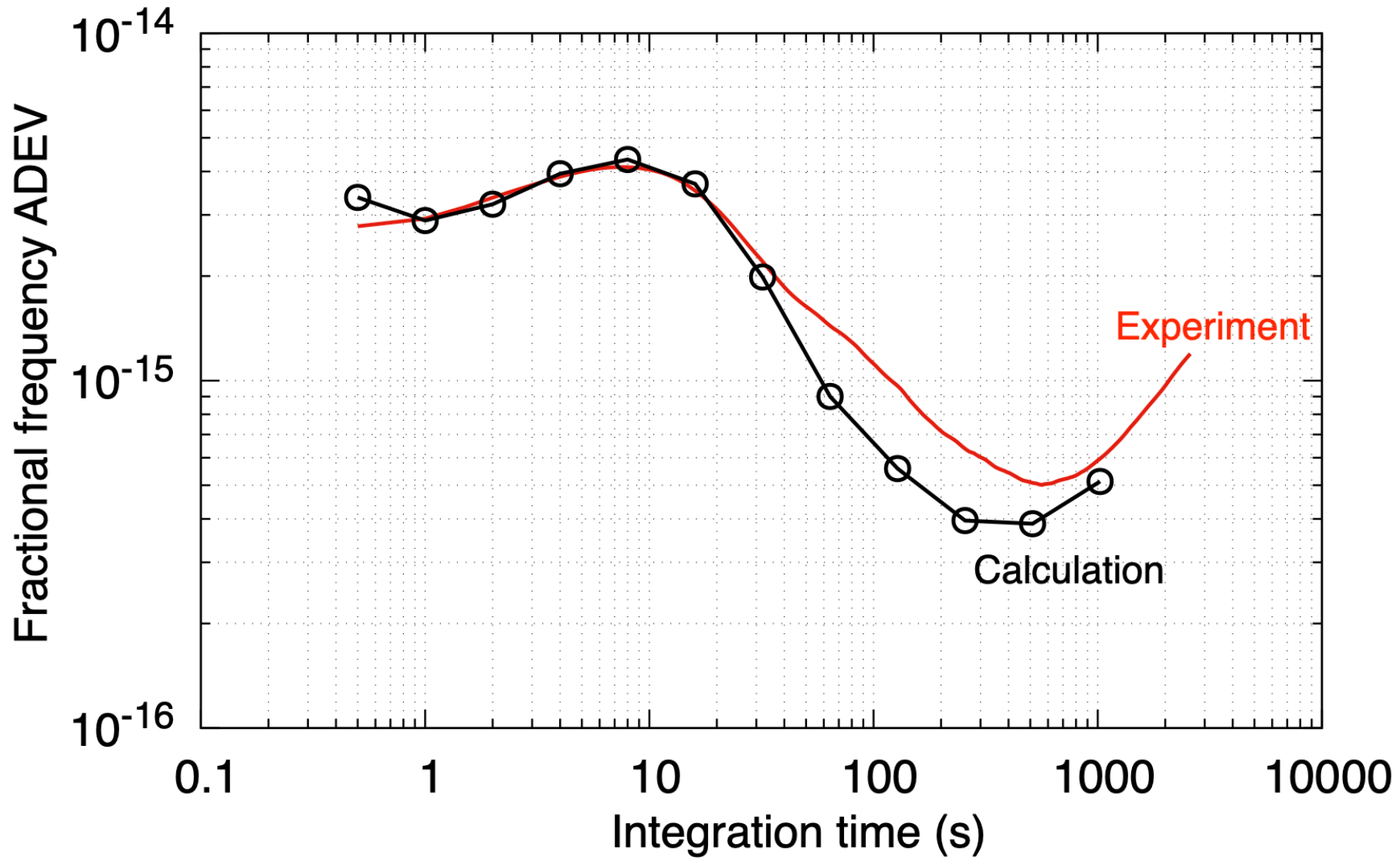}
\caption{\small \it CSO fractional frequency stability. Red line: experimental ADEV, Black line: calculated from the temperature spectrum.}
\label{fig-10}
\end{figure}

Here the controller parameters have been adjusted for a smooth control with a bandwidth of $\sim 0.02$ Hz. For $f~ \gtrsim~0.1$~Hz, the resonator temperature fluctuations are below the intrinsic noise of the temperature measurement carried out by the Cernox probe. We thus assume that the spectrum of the temperature fluctuations experienced by the resonator corresponds to the red line in the figure \ref{fig-9} with values to be read on the right scale. The resulting ADEV has been calculated and compared to the experimental result shown in the figure \ref{fig-10}.  

The experimental CSO frequency stability is well represented by the calculated ADEV, meaning that the delayed response of the paramagnetic ions is indeed responsible for the ADEV degradation.

\section{Summary}
We demonstrated that the dynamical thermal sensitivity imposed by the paramagnetic impurities degrades the CSO frequency stability. Solving this issue is not straightforward as the presence of these ions is essential to cancel the first order thermal sensitivity. 

The first solution to this issue is to filter out temperature fluctuations more effectively. Increasing the ballast mass is not possible without a complete redesign of the cryostat. We are instead thinking of integrating high-specific-heat materials such as HoCu$_2$ \cite{bao2015,zhi2021}.
Another solution is to find sapphire crystals with less chromium impurities. In the first crystals we purchased about 30 years ago, the dominant impurity was Mo$^{3+}$. This paramagnetic ion presents a high ESR frequency, i.e. $165$~GHz, and a short spin-lattice relaxation time of the order of $1$~ms \cite{sharoyan74}. This should permit to gain more than one order of magnitude on the $\tilde{a}$ parameter. Besides, the first CSOs prototypes we built with these crystals had frequency instabilities of less than $10^{-15}$~ (Ref. \onlinecite{journal-physics-2016}).  

\section*{Acknowledgements}
This work was supported by the LABEX Cluster of Excellence FIRST-TF (ANR-10-LABX-48-01), within the Program “Investissements d'Avenir” operated by the French National Research Agency (ANR).

The authors would like to thank Rodolphe Boudot and Enrico Rubiola for reviewing the manuscript and for their comments and corrections, which helped improve the original text.
\section*{DATA AVAILABILITY STATEMENT}
The data that support the findings of this study are available from the corresponding author upon reasonable request.

\end{document}